%% file: paper.tex
\documentclass[conference]{IEEEtran}
\IEEEoverridecommandlockouts

\usepackage{cite}
\usepackage{amsmath,amssymb,amsfonts}
\usepackage{algorithmic}
\usepackage{graphicx}
\usepackage{textcomp}
\usepackage{color}
\def\BibTeX{{\rm B\kern-.05em{\sc i\kern-.025em b}\kern-.08em
    T\kern-.1667em\lower.7ex\hbox{E}\kern-.125emX}}
    
\newcommand{\tool}{{Caprese}\xspace}
\newcommand{\tarmaq}{{TARMAQ}\xspace}

\newcommand{\js}{{Java\-Script}\xspace}
\newcommand{\numberOfSubjects}{{10}\xspace}

\usepackage{booktabs}

\usepackage{algorithmic}
\usepackage{algorithm}
\usepackage{listings}
\usepackage{color}
\usepackage{multirow}
\usepackage{array}
\usepackage{xspace}
\usepackage{hyperref}
\usepackage[utf8]{inputenc}
\usepackage{balance}
\usepackage{arydshln}
\usepackage{amsmath}
\usepackage{soul}  
\setulcolor{red}

\definecolor{gray}{rgb}{0.5,0.5,0.5}
\definecolor{mauve}{rgb}{0.58,0,0.82}
\definecolor{oliveGreen}{rgb}{0.5,.75,0.32}

\lstdefinelanguage{JavaScript}{
  morekeywords=[1]{break, continue, delete, else, for, function, if, in,
    new, return, this, typeof, var, void, while, with},
  morekeywords=[2]{false, null, true, boolean, number, undefined,
    Array, Boolean, Date, Math, Number, String, Object},
  morekeywords=[3]{eval, parseInt, parseFloat, escape, unescape},
  sensitive,
  morecomment=[s]{/*}{*/},
  morecomment=[l]//,
  morecomment=[s]{/**}{*/}, 
  morestring=[b]',
  morestring=[b]"
}[keywords, comments, strings]

\begin{document}

\title{Change Impact Recommendation for \js:\\Lessons from History and Runtime Analysis}

\author{\IEEEauthorblockN{Sadjad Tavakoli\IEEEauthorrefmark{1}}
\IEEEauthorblockA{
Magellan AI\\
Burnaby, BC, Canada \\
sadjad\_tavakoli@sfu.ca}
\thanks{\IEEEauthorrefmark{1} Work conducted while pursuing a graduate degree at Simon Fraser University.}

\and
\IEEEauthorblockN{Saba Alimadadi}
\IEEEauthorblockA{
Simon Fraser University\\
Burnaby, BC, Canada \\
saba@sfu.ca}
}

\maketitle

\input{Sections/Abstract}


\input{Sections/Introduction}
\input{Sections/Background}
\input{Sections/Approach}

\input{Sections/EvaluationSetup}
\input{Sections/EvaluationResults}

\input{Sections/Evaluation/threatsToValidity}
\input{Sections/RelatedWork}
\input{Sections/Conclusion}

\section*{Acknowledgements}
This work was supported in part by the Natural Sciences and Engineering Research Council of Canada (NSERC). The authors thank Hossein Soltanloo for his assistance with the experimental evaluation and the manual construction of the reference inspection sets.

\balance

\bibliographystyle{IEEEtran}
\bibliography{IEEEabrv,refs}

\end{document}

%% file: Sections/Abstract.tex
\begin{abstract}

Understanding the downstream effects of code changes is essential for software maintenance, debugging, and regression testing. This task is particularly challenging for \js applications, where dynamic language features such as callbacks, events, asynchronous execution, and shared mutable state make dependencies difficult to infer precisely.

Existing approaches for change impact recommendation rely primarily on either dependency-based analysis or mining software repositories. Dependency-based techniques, particularly dynamic analysis, can capture runtime interactions that reflect observed execution behaviour, but may miss relationships that are not exercised during analysis. In contrast, history-based techniques can uncover evolutionary coupling from past changes, but often introduce imprecise recommendations due to noisy co-change patterns.

To investigate the strengths and limitations of these approaches in \js, we engineer and evaluate three practical recommendation techniques: a history-based approach using co-change pattern mining, a dynamic dependency-based approach, and a hybrid approach that combines both signals. We implement these techniques in a unified framework, \tool, and evaluate them on \numberOfSubjects\ open-source Node.js applications using expert-curated reference inspection sets.

Our results reveal surprisingly low overlap between candidates identified by history-based and dynamic analyses, with only 22\% overlap at broader inspection budgets, indicating that the two approaches capture complementary notions of change impact. Dynamic analysis generally yields higher recommendation precision, while history-based analysis identifies additional relevant candidates missed by dependency analysis. We further show that combining historical and dynamic signals provides robust recommendation quality across diverse projects and inspection scenarios. These findings suggest that practical change impact recommendation in \js benefits from combining complementary runtime and evolutionary signals, as no single technique sufficiently captures all relevant inspection candidates.

\end{abstract}

%% file: Sections/Introduction.tex
\section{Introduction}
\label{sec:introduction}

Understanding the downstream effects of source code changes is essential during software development and maintenance. When modifying code, developers often need to determine which parts of the system may also require inspection, modification, retesting, or further validation. Such information can support important engineering tasks including debugging, regression testing, code review, and software evolution. Change impact recommendation aims to assist this process by identifying and prioritizing code entities likely to require inspection following a given change~\cite{theFirstImpactanalysis1996, Chianti2004, ImpactMiner2012, domSensitive2015}.

Providing accurate change impact recommendations is particularly challenging for highly dynamic languages such as \js. Unlike more statically analyzable languages, \js applications frequently rely on dynamic dispatch, first-class functions, callbacks, asynchronous execution, event-driven communication, timers, and shared mutable state. These language features often create dependencies that are implicit, runtime-dependent, or difficult to infer statically, making precise impact reasoning significantly harder~\cite{sabaUnderstanding2014, domSensitive2015, FrankTip2013}.

A common family of techniques for change impact recommendation relies on dependency analysis. Traditional static approaches use control-flow and data-flow relationships to estimate downstream effects of changes. However, static analysis often suffers from high imprecision, especially for dynamic languages such as \js. Dynamic analysis has therefore emerged as a practical alternative, as it can capture runtime interactions that are difficult to infer statically, including callback relationships, event-based communication, and asynchronous execution flows. However, dynamic analysis remains inherently incomplete because it only observes behaviours exercised during execution and may miss dependencies in unexecuted paths~\cite{Hattori2008, ramanathan2006sieve, heterogeneous2011, MesbahCIA2019, gyori2017refining}.

Another major family of techniques relies on mining software repositories. These history-based approaches exploit evolutionary coupling, based on the intuition that program entities that frequently change together in the past are likely to be related and may change together again in the future. Such techniques are attractive because they are largely language-agnostic and can uncover relationships that are not directly visible in the code, including semantic and maintenance coupling. However, historical co-change signals can be noisy: entities may co-change due to refactoring, tangled commits, or non-causal maintenance activities, resulting in imprecise recommendations~\cite{HassanHolt2004, Zimmermann2005, TARMAQ2016, Islam2018}.

These observations suggest that dependency-based and history-based techniques capture fundamentally different signals. Dependency analysis captures structural and runtime relationships, while repository mining captures evolutionary relationships embedded in software history. In practice, however, it remains unclear how these signals compare in the context of \js, what each technique misses, and whether combining them can improve recommendation quality.

In this paper, we report our experience engineering and evaluating three practical change impact recommendation approaches for \js: (1) a history-based approach that mines co-change patterns from software repositories, (2) a dynamic dependency-based approach that infers function-level runtime dependencies from execution traces, and (3) a hybrid approach that combines both signals. We implement these approaches in a unified framework, \tool, enabling a systematic comparison of their strengths, weaknesses, and complementarities.

We evaluate these approaches on \numberOfSubjects\ open-source Node.js applications using expert-curated reference inspection sets constructed by two independent investigators following predefined \js-specific dependency guidelines. Our study shows surprisingly low overlap between candidate functions identified by the history-based and dynamic approaches, indicating that they capture complementary notions of change impact. We further show that hybrid recommendations provide a robust balance between the higher precision of dynamic analysis and the broader coverage of history-based recommendations. Based on these findings, we derive practical lessons for engineering change impact recommendation tools for dynamic languages such as \js.

The main contributions of this work are as follows:

\begin{itemize}
\item We engineer and implement three practical function-level change impact recommendation approaches for \js: history-based, dynamic dependency-based, and hybrid approaches, and release the implementation and experimental artifacts to support reproducibility.~\footnote{\url{https://github.com/SEatSFU/caprese}}

\item We conduct an empirical evaluation of these approaches on \numberOfSubjects\ open-source Node.js applications using expert-curated reference inspection sets.

\item We provide practical insights into the strengths, limitations, and complementarities of historical and dynamic signals for change impact recommendation in highly dynamic software systems.

\end{itemize}

%% file: Sections/Background.tex
\section{Change Impact Recommendation for \js}
\label{sec:background}

\subsection{Change Impact Recommendation}

In this work, a \textit{change set} (CS) is defined as the set of functions modified in a given commit. Given a change set, the goal of \textit{change impact recommendation} is to identify additional functions likely to require inspection due to potential downstream effects of the change.

We model the output of a recommendation approach as a ranked \textit{Candidate Inspection Set} (CIS), where functions appearing earlier in the ranking are estimated to be more relevant for inspection. Ranking is important in practice because developers typically operate under limited inspection budgets and may only examine the top-$k$ recommended functions.

For evaluation, we use an expert-curated \textit{Reference Inspection Set} (RIS), representing a reference set of functions judged likely to require inspection for a given change set according to predefined dependency guidelines. The RIS serves as an evaluation reference rather than an exhaustive ground truth.

\subsection{Multiple Notions of Impact}

The notion of ``impact'' in modern software systems is inherently multi-dimensional. A code change may affect other functions through different kinds of relationships, and no single representation fully captures all such relationships.

\textit{Structural impact} arises from explicit program dependencies such as call relationships, control flow, and data flow. Traditional static change impact analysis primarily relies on these dependencies.

\textit{Runtime impact} arises from interactions observed during execution, including callback invocation, event propagation, asynchronous communication, and shared-state access. These dependencies are often difficult or impossible to infer precisely through static analysis alone.

\textit{Evolutionary impact} arises from historical co-change patterns in software repositories. Functions that frequently evolve together may reflect semantic or maintenance coupling even when explicit code dependencies are absent.

These complementary notions of impact motivate the three recommendation approaches studied in this paper.

\subsection{\js-Specific Challenges}
\label{subsec:js-challenges}

Change impact recommendation is particularly challenging in \js due to the dynamic and event-driven nature of the language and its ecosystem. Unlike more statically analyzable languages, dependencies between functions in \js are often implicit, runtime-dependent, and distributed across asynchronous execution paths.

First, \js heavily relies on \textit{callbacks}, \textit{promises}, and \textit{async/await} for asynchronous execution. As a result, a function may indirectly trigger the execution of other functions without explicit call relationships visible in local control flow. For example, a function that registers a callback may affect downstream behaviour long after the original function returns.

Second, many \js applications, and particularly Node.js systems, use event-driven communication through mechanisms such as \texttt{EventEmitter}. In such systems, a function may emit an event whose handlers are registered elsewhere in the codebase, creating dependencies that are difficult to capture using conventional call-graph analysis.

Third, deferred execution mechanisms such as \texttt{setTimeout}, task queues, and asynchronous I/O further complicate dependency reasoning by separating cause and effect across time. A code change may therefore affect functions that execute much later and in a different execution context.
%

Finally, shared mutable state through objects, closures, and module-level variables can introduce implicit data dependencies between functions even when no explicit call relationship exists. In addition, some related functions may exhibit no observable runtime dependency at all, yet still evolve together historically due to architectural conventions, duplicated logic, or maintenance practices.

These characteristics make \js particularly challenging for change impact recommendation and motivate the need for multiple complementary signals. No single signal or analysis technique can capture all relevant dependencies reliably, which motivates the repository-based, dynamic, and hybrid approaches studied in this paper.

The next section describes how we operationalize these complementary signals into repository-based, dynamic, and hybrid recommendation approaches.









%% file: Sections/Approach.tex
\section{Engineering the Three Recommendation Approaches}
\label{sec:approach}

\autoref{fig:overview} presents an overview of our framework, \tool. Given a change set, the framework generates a \textit{Candidate Inspection Set (CIS)} using three approaches: a repository-based approach that mines historical co-change patterns, a dynamic dependency-based approach that infers runtime dependencies from execution traces, and a hybrid approach that combines both signals. Each approach produces a ranked Candidate Inspection Set (CIS), enabling direct comparison of recommendation quality and signal complementarity.

\subsection{Repository-Based Recommendation}
\label{subsec:history}


The repository-based approach relies on the intuition of \textit{evolutionary coupling}: if two functions repeatedly changed together in the past, a future change in one function may warrant inspection of the other. Historical co-change can reveal semantic and maintenance relationships that may not be explicitly visible in source code, making repository mining a useful signal for change impact recommendation.

Given a change set, this approach recommends functions that have frequently co-changed with one or more functions in the change set throughout the project’s history. The resulting recommendations form a Candidate Inspection Set (CIS) ranked according to the strength of the observed co-change patterns.

\subsubsection{Repository Mining Strategy}

Our implementation is inspired by prior co-change recommendation techniques such as \tarmaq~\cite{TARMAQ2016}, but is adapted and extended for function-level recommendation in \js systems. Unlike traditional repository-based approaches that operate at file granularity and primarily model pairwise co-change relationships, our approach performs targeted function-level pattern mining to discover higher-order co-change relationships relevant to the current change set.

We first extract the set of changed functions from each commit in the project history, treating each commit as a \textit{transaction} consisting of all functions modified together. To identify changed functions at function granularity, we compare consecutive revisions using RefDiff’s~\footnote{\url{https://github.com/aserg-ufmg/RefDiff}} code structure trees (CSTs), which allow mapping changed lines to their enclosing functions, classes, and files. This enables language-aware extraction of function-level change history for \js projects.

To mine relevant co-change patterns, we use a targeted pattern expansion procedure inspired by frequent and pattern mining algorithms. Starting from singleton patterns containing functions from the change set, candidate patterns are iteratively expanded by exploring co-occurring functions in relevant transactions. Candidate patterns whose support falls below a minimum threshold are pruned early to reduce search cost and control combinatorial explosion. This targeted strategy avoids mining all repository-wide patterns upfront while focusing the search on patterns relevant to the current recommendation task.

To reduce noise, we exclude commits that are unlikely to provide meaningful co-change information: (1) commits modifying only a single function, since they contain no co-change information; (2) merge commits, which often duplicate changes from multiple branches; and (3) large commits affecting more than 30 files, which typically correspond to housekeeping, formatting, or large-scale refactoring activities rather than meaningful functional coupling.

For each discovered pattern $P$, we compute its \textit{support}, defined as the number of transactions containing all functions in $P$, and its \textit{confidence}, defined as:

\begin{figure}

\includegraphics[clip, trim=0cm 16.5cm 16.5cm 0cm, width=0.5\textwidth]{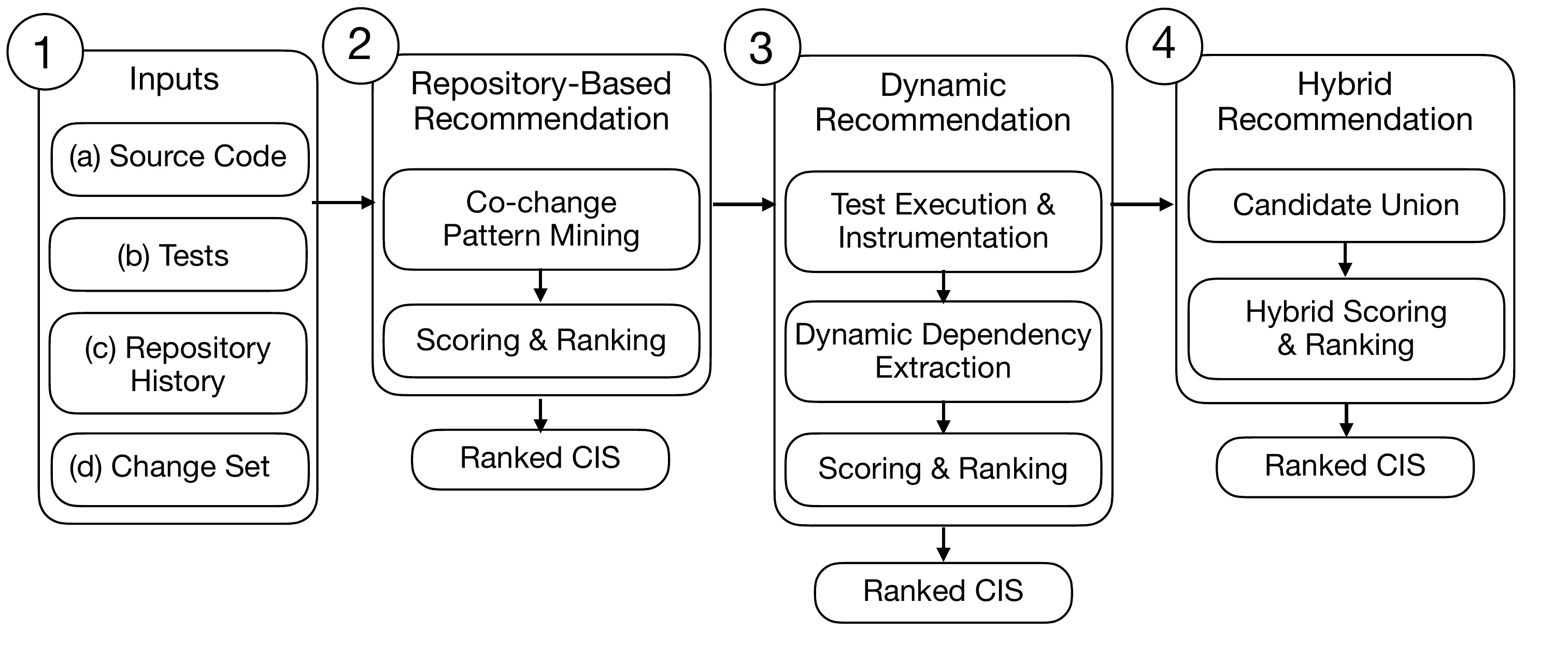}

\caption{Overview of the recommendation workflow implemented in \tool. (1) The framework receives the change set together with project artifacts (source code, tests, repository history, and change set) as input. (2) The repository-based approach mines historical co-change patterns to generate a ranked Candidate Inspection Set (CIS). (3) The dynamic approach extracts runtime dependencies from instrumented executions and generates a ranked CIS. (4) The hybrid approach combines candidates and ranking signals from both approaches to produce a final ranked CIS.}

\label{fig:overview}
\end{figure}

\begin{equation}
\label{eq:confidence}
confidence(P)=\frac{support(P)}{support(P_c)}
\end{equation}

where $P_c$ denotes the subset of pattern functions belonging to the change set. Support captures how consistently a pattern appears in project history, while confidence measures how strongly the remaining pattern functions are associated with the change set. High-support but low-confidence patterns represent common yet weakly predictive co-change, whereas high-confidence patterns provide stronger recommendation evidence. Patterns whose confidence exceeds a predefined threshold contribute candidate functions to the CIS.

\subsubsection{Practical Engineering Adaptations}

Compared with prior repository-based approaches, our implementation introduces several engineering adaptations specifically designed for function-level change impact recommendation in \js.

\textbf{Function-level granularity.}
Recommendations are produced at the function level rather than file level, providing more actionable inspection candidates.

\textbf{\js-aware function extraction.}
Using CST-based extraction enables accurate mapping from commits to changed \js functions despite language dynamism and refactoring.

\textbf{Pattern expansion beyond pairwise co-change.}
Instead of considering only pairwise co-change relationships, we mine higher-order patterns involving multiple functions, allowing the discovery of richer co-evolution structures.

\textbf{Deleted-function pruning.}
Because software evolves over time, historical patterns may include functions that no longer exist. We explicitly track deleted functions and remove them from final recommendations.

For example, consider a change set containing \texttt{storeData}. Suppose the project history contains commits in which \texttt{storeData} changed together with \texttt{validateGradsUsernames}, \texttt{validateProfsUsernames}, and \texttt{validateGradsEmails}. These functions may not all be directly connected through explicit calls, but they operate over related user data and persisted files. A repository-based approach can therefore surface them as inspection candidates because the history indicates that developers have maintained them together. In this setting, support captures how often such functions co-change, while confidence estimates how strongly the candidate functions are associated with the changed functions in the current change set.

\subsubsection{Strengths and Limitations}

A key strength of repository-based recommendation is that it is largely language-agnostic and can uncover semantic or maintenance coupling even when explicit code dependencies are absent. However, historical co-change signals can be noisy: functions may co-change due to refactoring, tangled commits, or other non-causal maintenance activities. As a result, co-change does not necessarily imply true dependency, which can reduce recommendation precision.

\subsection{Dynamic Dependency Recommendation}
\label{subsec:dynamic}


The dynamic dependency-based approach relies on the intuition that if a function interacts with another function during execution, then modifying one may require inspection of the other. Unlike repository-based recommendation, which infers relationships from historical co-change, this approach derives dependencies directly from observed runtime behaviour. This enables the detection of execution-based relationships that are often difficult to infer from source code alone.

Given a change set, this approach recommends functions that are reachable through runtime dependencies originating from the changed functions. These recommendations form a Candidate Inspection Set (CIS) ranked according to dependency strength and proximity in the dependency graph.


We do not claim that static analysis is ineffective for change impact recommendation. However, due to \js's highly dynamic runtime behaviour, purely static dependency inference can be difficult and often imprecise. Features such as dynamic dispatch, first-class functions, callbacks, event-driven communication, asynchronous execution, and shared mutable state frequently create dependencies that are implicit or only observable at runtime~\cite{domSensitive2015}.
For this reason, we use dynamic dependency extraction as a practical approximation of runtime interactions. By observing actual executions, dynamic analysis can capture relationships that static analysis may miss, particularly in asynchronous and event-driven Node.js applications.

\subsubsection{Dependency Graph Construction}

We construct a dynamic dependency graph by instrumenting the target application and collecting execution traces during test execution. Our implementation uses NodeProf,~\footnote{\url{https://www.dag.inf.usi.ch/software/nodeprof}}
a dynamic analysis framework for Node.js,
to observe runtime interactions among program functions and their associated execution contexts.

The resulting dependency graph contains nodes representing program entities and edges representing observed dependencies between them. We capture several dependency types relevant to \js applications:

\textbf{Function call dependencies.}
An edge is created when one function directly invokes another.

\textbf{Callback dependencies.}
Dependencies are recorded when functions are passed and later invoked as callbacks, including asynchronous callbacks.

\textbf{Event-based dependencies.}
Dependencies are created between event emitters and registered event handlers, capturing common event-driven communication patterns.

\textbf{Timer-based dependencies.}
Deferred execution through mechanisms such as \texttt{setTimeout} and scheduling queues creates dependencies across time-separated executions.

\textbf{Shared-state dependencies.}
Read and write operations on shared objects, closures, or module-level variables create implicit data dependencies between otherwise disconnected functions.

These dependency types allow the graph to capture both explicit and implicit runtime relationships that are especially important in \js systems.

As an example, consider a function that emits an event and another function that registers a listener for that event (\autoref{fig:dynamic-example}, line 1).
A simple local call graph may not directly connect the emitter to the listener, yet at runtime the emitted event triggers the listener callback (line 2). If that callback later schedules another function through \texttt{setTimeout} (line 3), and the scheduled function invokes additional functions (line 4), the impact of the original change may propagate through an event-driven and time-delayed execution chain. Dynamic dependency extraction is designed to capture these runtime relationships and represent them as edges in the dependency graph.

\begin{figure}[t]
\centering
\small
\fbox{
\begin{minipage}{0.9\columnwidth}
1. \texttt{foo() $\rightarrow$ emit("event")}\\
2. \texttt{bar() $\rightarrow$ on("event", baz)}\\
3. \texttt{baz() $\rightarrow$ setTimeout(qux)}\\
4. \texttt{qux() $\rightarrow$ fred(); foz();}
\end{minipage}
}
\caption{Example of implicit runtime dependencies in \js. Although \texttt{foo} does not directly call \texttt{baz} or \texttt{qux}, event emission and deferred callbacks create dependency chains that are difficult to infer statically.}
\label{fig:dynamic-example}
\end{figure}

\subsubsection{Candidate Generation}

Given a change set, we perform bounded graph traversal starting from changed-function nodes to collect reachable candidate functions. Functions closer to the changed nodes in the graph are generally considered stronger candidates for inspection than those farther away.
Traversal proceeds outward from change-set nodes following dependency edges up to a bounded depth. During traversal, we accumulate evidence from multiple dependency paths and edge types. Candidates reachable through multiple short paths or stronger dependency edges receive higher ranking scores than those reachable only through longer or weaker paths. This prioritizes functions that are more directly influenced by the change set while limiting noise from distant transitive dependencies.

The resulting Candidate Inspection Set is ranked using graph-based features such as dependency distance and connection strength, prioritizing functions with stronger or more direct runtime relationships to the change set.

\subsubsection{Strengths and Limitations}

A key strength of dynamic dependency recommendation is its ability to capture precise runtime relationships, including asynchronous and event-driven interactions that are difficult to infer statically. However, its effectiveness depends heavily on execution coverage. Dependencies that are not exercised during tracing remain invisible, which can lead to incomplete recommendations.

\subsection{Hybrid Recommendation}
\label{subsec:hybrid}

\subsubsection{Motivation}

The repository-based and dynamic dependency-based approaches capture complementary signals for change impact recommendation. Repository mining can uncover semantic and maintenance coupling that may not be observable during execution, while dynamic analysis captures precise runtime interactions that are often absent from historical co-change patterns. Since each approach reveals relationships missed by the other, combining both signals can potentially improve recommendation quality.

\subsubsection{Combination Strategy}

Our hybrid approach combines the Candidate Inspection Sets (CIS) produced by the repository-based and dynamic approaches into a unified ranked recommendation list. The combined candidate set is formed by taking the union of functions identified by both approaches, followed by filtering invalid candidates such as obsolete functions.

Candidate ranking integrates information from both recommendation signals. Repository-based candidates contribute pattern-based features such as co-change support and confidence, while dynamic candidates contribute graph-based features such as dependency strength and distance from the change set.

In practice, ranking prioritizes candidates based on both the strength and agreement of the two signals. Functions identified by both approaches receive the highest priority, as agreement between historical and runtime evidence provides stronger confidence in their relevance. Among repository-based candidates, higher support and confidence increase priority, reflecting stronger historical co-change evidence. Among dynamic candidates, functions with shorter dependency distance and stronger connection strength to the change set are ranked higher. Intuitively, functions that repeatedly co-change with the change set and also exhibit strong runtime dependencies are more likely to represent high-value inspection candidates.

\subsubsection{Expected Benefits}

The hybrid approach aims to leverage the strengths of both recommendation strategies while mitigating their individual weaknesses. Historical signals can recover relevant candidates missed due to incomplete execution coverage, whereas dynamic signals can reduce noise from non-causal co-change patterns. As a result, we expect the hybrid approach to improve recommendation precision while maintaining broader candidate coverage than either individual approach alone.

%% file: Sections/EvaluationSetup.tex
\section{Evaluation Setup}
\label{sec:evaluation}

This section describes the methodology used to evaluate the three change impact recommendation approaches presented in \autoref{sec:approach}. Our evaluation focuses on understanding the complementarity of historical and runtime signals, the effect of hybrid recommendation on recommendation quality, and the practical lessons that emerge from applying these approaches to real-world \js systems.

\subsection{Research Questions}

Our evaluation is guided by the following research questions:

\textit{\textbf{RQ1:} How different are the recommendations from repository-based and dynamic analyses?}

This question evaluates the degree of overlap and complementarity between historical and runtime recommendation signals. In particular, we investigate how many relevant candidate functions are uniquely identified by each approach versus jointly identified by both.

\textit{\textbf{RQ2:} How does hybrid recommendation affect recommendation quality?}

This question evaluates whether combining repository-based and dynamic signals improves recommendation quality compared to using either signal individually. We compare recommendation precision across repository-based, dynamic, hybrid, and TARMAQ-based approaches.

\textit{\textbf{RQ3:} What practical lessons emerge from applying these approaches?}

This question focuses on qualitative analysis of failure modes, error patterns, and engineering tradeoffs observed during evaluation.

\subsection{Subject Applications}

We evaluate our approaches on \numberOfSubjects open-source Node.js applications spanning diverse sizes and domains, including web frameworks, middleware libraries, development tools, and utility systems. The benchmarks range from 4 to 49 KLOC and from 220 to 5,248 commits, providing substantial variation in both codebase size and repository history. All selected projects include executable test suites, enabling dynamic dependency extraction through runtime tracing.

\autoref{tab:benchmarks} summarizes the benchmark characteristics.

\subsection{Reference Inspection Set Construction}

Evaluating change impact recommendation requires a reference set of functions considered relevant for inspection following a code change. Since exhaustive ground truth is difficult to establish for function-level impact in highly dynamic systems, we use expert-curated \textit{Reference Inspection Sets} (RIS) rather than assuming complete ground truth.

For each subject application, we randomly selected five commits from the project history as representative change sets, resulting in 50 evaluated change sets in total. We selected commits that modified a unique set of functions to avoid duplicate evaluation instances and ensure diversity in the analyzed changes. The remaining commits constituted the historical change data used by the repository-based recommendation approach.

For each selected change set, the corresponding RIS was manually constructed by two independent investigators, one of whom is not an author of this paper. The investigators first identified the functions modified by the selected commit and then examined additional functions that might require inspection due to potential downstream impact.

The investigators followed predefined \js-specific dependency guidelines and considered several dependency categories, including direct function calls, callback relationships, event-based interactions, shared-state dependencies, and semantic coupling requiring coordinated inspection. Functions deemed relevant for inspection were included in the RIS. Disagreements were resolved through discussion until consensus was reached.

\subsection{Evaluation Metrics and Inspection Budgets}

Each approach produces a ranked Candidate Inspection Set (CIS). In practice, developers rarely inspect all recommended functions and instead focus on top-ranked candidates under limited time and attention budgets.

We therefore evaluate recommendation quality using three \textit{inspection budgets} corresponding to cutoff points:

$K \in {5,20,60}$.

A cutoff point of $K$ means that only the top-$K$ functions in the ranked CIS are considered. These cutoffs simulate realistic inspection scenarios ranging from highly constrained review (Top-5) to broader maintenance-oriented inspection (Top-60). Smaller cutoff points place greater emphasis on ranking quality, as more relevant candidates are expected to appear earlier in the recommendation list.

\input{Sections/Evaluation/tablesData/benchmarksTable}

\input{tables/complementarity}

We use \textbf{Precision@K} as the primary evaluation metric:

[
Precision@K =
$\frac{\text{Relevant functions in top-}K}{K}$
]

where relevant functions are those appearing in the Reference Inspection Set (RIS).

For RQ1, we additionally measure recommendation complementarity by computing the proportion of relevant functions identified exclusively by repository-based analysis, exclusively by dynamic analysis, and by both approaches.

\subsection{Experimental Setup}

For repository-based recommendation, we mine historical commits after filtering merge commits, single-function commits, and large housekeeping commits affecting more than 30 files. The remaining commits are transformed into function-level change transactions used for co-change pattern mining.

For dynamic recommendation, we instrument each application using NodeProf~\footnote{\url{https://github.com/Haiyang-Sun/nodeprof.js/}}
 and execute its available test suite to collect runtime traces. These traces are used to construct function-level dependency graphs capturing explicit and implicit runtime relationships.

All approaches produce ranked Candidate Inspection Sets, enabling direct comparison of recommendation quality across identical benchmarks and inspection budgets.

%% file: Sections/Evaluation/tablesData/benchmarksTable.tex
\begin{table}[t]
\centering
\caption{Characteristics of the benchmark applications.}
\label{tab:benchmarks}
\small
\renewcommand{\arraystretch}{1.1}
\begin{tabular}{lrrrr}
\toprule
\textbf{Project} & \textbf{KLOC} & \textbf{Commits} & \textbf{Transactions} & \textbf{Functions} \\
\midrule
bignumber.js   & 48 &   220 &   111 &   347 \\
session        &  4 &   678 &   257 &   723 \\
jhipster-uml   & 11 &   868 &   448 & 2,676 \\
grant          & 11 & 1,134 &   567 & 2,643 \\
environment    & 33 & 1,165 &   661 & 3,221 \\
cla-assistant  & 46 & 1,177 &   634 & 3,186 \\
assemble       & 26 & 1,662 &   926 & 4,465 \\
nock           & 36 & 1,756 &   882 & 5,847 \\
fastify        & 49 & 2,682 & 1,282 & 6,844 \\
express        & 19 & 5,248 & 3,111 & 9,179 \\
\midrule
\textbf{Total}   & \textbf{283}  & \textbf{16,590}  & \textbf{8,879} & \textbf{39,131} \\
\bottomrule
\end{tabular}
\end{table}

%% file: tables/complementarity.tex
\begin{table*}[t]
\centering
\caption{Complementarity of repository-based and dynamic recommendations across three inspection budgets. Values show the percentage of relevant recommendations identified exclusively by the repository-based approach, exclusively by the dynamic approach, or by both (overlap), with respect to the expert-curated reference inspection sets.}
\label{tab:signalComplementarity}
\small
\renewcommand{\arraystretch}{1.1}
\begin{tabular}{lrrr|rrr|rrr}
\toprule
\multirow{2}{*}{\textbf{Project}} &
\multicolumn{3}{c|}{\textbf{Top-5}} &
\multicolumn{3}{c|}{\textbf{Top-20}} &
\multicolumn{3}{c}{\textbf{Top-60}} \\
\cmidrule(lr){2-4}
\cmidrule(lr){5-7}
\cmidrule(lr){8-10}
&
\textbf{Repo} &
\textbf{Dynamic} &
\textbf{Overlap} &
\textbf{Repo} &
\textbf{Dynamic} &
\textbf{Overlap} &
\textbf{Repo} &
\textbf{Dynamic} &
\textbf{Overlap} \\
\midrule
bignumber.js   & 32.5 &  2.5 & 65.0 & 35.2 &  3.8 & 61.0 & 61.8 & 19.4 & 18.9 \\
session        & 11.5 & 12.0 & 76.5 & 29.7 & 16.1 & 54.3 & 44.5 & 20.9 & 34.6 \\
jhipster-uml   & 24.1 & 14.8 & 61.1 & 39.0 & 32.3 & 28.7 & 49.3 & 31.4 & 19.3 \\
grant          & 15.6 & 56.3 & 28.1 & 28.6 & 59.4 & 11.1 & 26.4 & 65.2 &  7.7 \\
environment    & 24.8 &  7.8 & 67.4 & 27.7 & 13.6 & 58.7 & 51.1 & 17.1 & 31.7 \\
cla-assistant  &  5.6 & 44.4 & 50.0 & 18.8 & 54.8 & 26.4 & 19.9 & 60.4 & 19.7 \\
assemble       & 20.4 & 72.2 &  7.4 & 28.9 & 68.7 &  2.5 & 34.2 & 64.3 &  1.5 \\
nock           & 22.5 &  0.0 & 77.5 & 38.2 &  8.0 & 53.8 & 53.2 &  8.8 & 38.0 \\
fastify        &  0.0 & 14.3 & 85.7 & 20.0 & 27.9 & 45.5 & 47.1 & 21.0 & 24.6 \\
express        &  7.1 & 35.7 & 57.1 & 20.3 & 45.3 & 34.4 & 29.6 & 46.6 & 23.8 \\
\midrule
\textbf{Average}
& \textbf{16.4} & \textbf{26.0} & \textbf{57.6} 
& \textbf{28.6} & \textbf{33.0} & \textbf{37.6} 
& \textbf{41.7} & \textbf{35.5} & \textbf{22.0} \\
\bottomrule
\end{tabular}
\end{table*}

%% file: Sections/EvaluationResults.tex
\section{Results and Discussion}
\label{sec:results}

\subsection{RQ1: How different are repository-based and dynamic recommendations?}

\autoref{tab:signalComplementarity} summarizes the overlap between correctly recommended functions produced by the repository-based and dynamic approaches across three inspection budgets. The results show substantial complementarity between the two signals. At Top-5, the two approaches agree on a majority of relevant recommendations, with an average overlap of 57.6\%. However, as the inspection budget increases, the overlap decreases substantially. At Top-60, only 22.0\% of relevant recommendations are identified by both approaches, while 41.7\% are identified exclusively by the repository-based approach and 35.5\% exclusively by the dynamic approach.

This pattern is also consistent with the hybrid ranking strategy, which prioritizes candidates supported by both signals. As a result, functions identified by both repository-based and dynamic analyses tend to appear near the top of the ranked inspection set, leading to higher overlap at smaller cutoff points. As the inspection budget increases, more uniquely identified candidates from each approach become visible, causing the overlap to decrease.

This trend indicates that repository-based and dynamic analyses capture fundamentally different notions of change impact. Dynamic analysis identifies runtime relationships exercised during execution, while repository mining captures evolutionary relationships embedded in project history. The low overlap at larger inspection budgets suggests that the two approaches continue to contribute distinct relevant candidates beyond the highest-confidence recommendations.

We also observe notable project-level variation. For example, projects such as \textit{assemble} and \textit{grant} are heavily dominated by dynamic-only recommendations, suggesting stronger runtime coupling, whereas projects such as \textit{nock} and \textit{bignumber.js} exhibit larger repository-only contributions at broader inspection budgets. This variability further reinforces that the relative usefulness of recommendation signals depends on project characteristics and architecture.

These results indicate that repository-based and dynamic recommendation should be viewed as complementary rather than competing strategies, as each uncovers relevant inspection candidates that the other systematically misses.

\input{tables/precision}

\subsection{RQ2: How does hybrid recommendation affect recommendation quality?}

\autoref{tab:precisionComparison} reports Precision@K across all subject applications. Dynamic recommendation achieves the highest precision under small and medium inspection budgets, with average Precision@5 of 0.71 and Precision@20 of 0.51. This suggests that runtime dependencies provide a strong signal for prioritizing a small number of highly relevant inspection candidates.

Across all approaches, precision generally decreases as the inspection budget increases. This trend indicates that the ranking mechanisms are effective in prioritizing more relevant candidates near the top of the recommendation list. In practice, this is important because developers rarely inspect large candidate sets exhaustively; therefore, placing the most relevant candidates earlier substantially improves the practical usefulness of the recommendations.

The repository-based approach is less precise at small cutoffs but remains comparatively stable as the inspection budget grows, achieving the highest precision at Top-60 (0.41). This reflects the broader but noisier nature of historical co-change information. Compared with TARMAQ, the repository-based approach improves precision across all inspection budgets, suggesting that function-level co-change pattern mining provides more useful recommendations than pairwise evolutionary coupling alone.

The hybrid approach consistently achieves strong precision across all inspection budgets while outperforming TARMAQ and remaining competitive with the strongest individual approach at each cutoff point. This suggests that hybrid recommendation is particularly useful in realistic inspection settings where developers benefit from both highly precise runtime-based candidates and additional historically relevant candidates that may otherwise be missed. While hybrid recommendation does not always achieve the highest average precision, it provides more robust performance across diverse projects and inspection scenarios.

We also observe substantial project-level variation in recommendation quality. For example, dynamic analysis performs particularly well on projects such as \textit{bignumber.js}, \textit{cla-assistant}, and \textit{nock}, where runtime interactions appear to be strongly captured by the available execution traces. In contrast, repository-based recommendation performs comparatively better on projects such as \textit{express} and \textit{assemble}, suggesting stronger historical or semantic coupling not fully reflected in observed runtime dependencies. These differences further indicate that project architecture, development practices, and test coverage influence which recommendation signal is most effective.

Overall, the results indicate that no single recommendation strategy dominates across all inspection scenarios. Dynamic analysis is most effective when representative execution traces are available and developers can inspect only a small number of candidates. Repository-based recommendation becomes more useful under broader inspection budgets or when historical co-change reveals dependencies not captured dynamically. Hybrid recommendation provides a robust compromise by combining both signals and reducing dependence on any single source of evidence.

\subsection{RQ3: What practical lessons emerge from applying these approaches?}

Our manual analysis of the recommendation results revealed several practical lessons regarding the strengths and limitations of repository-based, dynamic, and hybrid recommendation.

Repository-based recommendation was particularly useful for identifying relevant candidates that were not visible through runtime dependency analysis. We observed four recurring categories of such useful repository-only discoveries. First, co-change patterns often revealed \textit{semantic coupling} between functions with similar purpose or structure, including duplicated or near-duplicated logic, even when no explicit runtime dependency existed between them. Second, repository history exposed dependencies mediated through \textit{external resources} such as APIs, databases, files, and caches, where impact propagation occurred outside the scope of user-written code. Third, co-change analysis captured \textit{primitive and configuration coupling}, including shared file paths, URLs, event names, error codes, request headers, and other string-based identifiers commonly used in \js applications. Finally, repository-based recommendation surfaced relevant candidates in \textit{unexecuted code paths}, including functions that would likely have been detected by dynamic analysis had they been exercised by the available tests. These observations suggest that repository mining can reveal semantic and maintenance coupling that dependency analysis alone may miss.

At the same time, repository-based recommendation introduced several recurring sources of noise. The most common source was \textit{tangled commits}, where unrelated changes were committed together, creating incidental co-change relationships that did not correspond to meaningful dependencies. We also observed \textit{deprecated dependencies}, where functions that historically changed together no longer exhibited relevant relationships in recent versions of the software. Another source of noise arose from \textit{asymmetric dependencies}: while co-change relationships are inherently bidirectional, real impact propagation is often directional, causing history-based methods to recommend functions in both directions even when only one direction is meaningful. Finally, we observed many cases where functions were related at a conceptual level but did not actually transfer impact, such as relationships between tests and the code under test. These findings reinforce that co-change is a useful but imperfect proxy for change impact.

Dynamic recommendation produced more precise top-ranked candidates, but its primary limitation was incomplete execution coverage. Dependencies not exercised by the available test suites remained invisible to the dynamic analysis. This limitation is particularly important in \js systems, where asynchronous callbacks, event handlers, timers, and framework-mediated behaviour may execute only under specific runtime conditions. As a result, dynamic analysis can miss relevant dependencies despite its strong precision for observed execution paths.

Hybrid recommendation inherits limitations from both approaches and may additionally suffer from ranking conflicts when historical and runtime signals disagree. In some cases, relevant candidates identified strongly by one signal may be ranked lower due to weaker evidence from the other signal. Nevertheless, combining both signals provides a useful balance between high-precision runtime recommendations and broader historical coverage.

Overall, these findings suggest that hybrid recommendation is most useful as \textit{inspection support} rather than exhaustive impact prediction. Change impact recommendation in highly dynamic systems such as \js is inherently multi-signal, and practical tools benefit most from exposing complementary evidence rather than attempting to predict a single complete impact set.

%% file: tables/precision.tex
\begin{table*}[t]
\centering
\caption{Precision@K of the recommendation approaches across three inspection budgets.}
\label{tab:precisionComparison}
\fontsize{7.8}{10}\selectfont
\renewcommand{\arraystretch}{1.1}
\begin{tabular}{lrrrr|rrrr|rrrr}
\toprule
\multirow{2}{*}{\textbf{Project}} &
\multicolumn{4}{c|}{\textbf{Top-5}} &
\multicolumn{4}{c|}{\textbf{Top-20}} &
\multicolumn{4}{c}{\textbf{Top-60}} \\
\cmidrule(lr){2-5}
\cmidrule(lr){6-9}
\cmidrule(lr){10-13}
&
\textbf{TARMAQ} &
\textbf{Repo} &
\textbf{Dynamic} &
\textbf{Hybrid} &
\textbf{TARMAQ} &
\textbf{Repo} &
\textbf{Dynamic} &
\textbf{Hybrid} &
\textbf{TARMAQ} &
\textbf{Repo} &
\textbf{Dynamic} &
\textbf{Hybrid} \\
\midrule
bignumber.js   & 0.39 & 0.60 & 0.92 & 0.76 & 0.41 & 0.44 & 0.88 & 0.68 & 0.33 & 0.48 & 0.57 & 0.59 \\
session        & 0.56 & 0.69 & 0.89 & 0.76 & 0.38 & 0.60 & 0.59 & 0.61 & 0.30 & 0.58 & 0.30 & 0.37 \\
jhipster-uml   & 0.22 & 0.33 & 0.55 & 0.34 & 0.14 & 0.25 & 0.26 & 0.26 & 0.07 & 0.25 & 0.12 & 0.14 \\
grant          & 0.36 & 0.51 & 0.75 & 0.46 & 0.21 & 0.44 & 0.60 & 0.41 & 0.17 & 0.45 & 0.33 & 0.29 \\
environment    & 0.20 & 0.36 & 0.67 & 0.62 & 0.26 & 0.36 & 0.56 & 0.43 & 0.28 & 0.31 & 0.37 & 0.36 \\
cla-assistant  & 0.30 & 0.36 & 0.80 & 0.48 & 0.21 & 0.33 & 0.43 & 0.37 & 0.19 & 0.31 & 0.24 & 0.27 \\
assemble       & 0.40 & 0.68 & 0.65 & 0.71 & 0.24 & 0.64 & 0.46 & 0.66 & 0.22 & 0.64 & 0.47 & 0.64 \\
nock           & 0.38 & 0.53 & 0.82 & 0.70 & 0.34 & 0.52 & 0.59 & 0.61 & 0.30 & 0.50 & 0.31 & 0.31 \\
fastify        & 0.14 & 0.35 & 0.33 & 0.36 & 0.11 & 0.18 & 0.23 & 0.20 & 0.09 & 0.16 & 0.19 & 0.17 \\
express        & 0.38 & 0.73 & 0.49 & 0.44 & 0.23 & 0.52 & 0.29 & 0.34 & 0.18 & 0.45 & 0.12 & 0.25 \\
\midrule
\textbf{Average}
& \textbf{0.33} & \textbf{0.49} & \textbf{0.71} & \textbf{0.58}
& \textbf{0.26} & \textbf{0.42} & \textbf{0.51} & \textbf{0.47}
& \textbf{0.22} & \textbf{0.41} & \textbf{0.32} & \textbf{0.35} \\
\bottomrule
\end{tabular}
\end{table*}

%% file: Sections/Evaluation/threatsToValidity.tex
\section{Threats to Validity}
\label{sec:threats}

An important \textbf{external threat} concerns the generalizability of our findings to all \js systems. To mitigate this threat, we evaluated our approaches on \numberOfSubjects open-source Node.js applications spanning diverse domains, codebase sizes, repository histories, and architectural styles. However, the selected benchmarks may not fully represent all \js ecosystems, particularly large industrial systems or applications using frameworks not covered in our dataset. Another external threat concerns the representativeness of the selected change sets. To reduce this threat, we evaluated multiple randomly selected change sets per project.

A key \textbf{construct threat} concerns the definition of recommendation quality. Since exhaustive ground truth for function-level change impact is difficult to establish in highly dynamic systems, we evaluated recommendation quality using expert-curated Reference Inspection Sets (RIS) rather than complete ground truth. As a result, our precision and coverage metrics reflect agreement with the RIS rather than absolute correctness. While this limitation is common in change impact analysis studies, it may affect the interpretation of recommendation quality.

The main \textbf{internal threat} concerns potential investigator bias during manual RIS construction due to the subjective nature of dependency analysis. Another threat arises from limited familiarity with the benchmark projects, since project maintainers might identify different inspection candidates. To reduce these threats, RIS construction was performed by two independent investigators, one of whom is not an author of this paper. The investigators had no prior involvement with the selected projects and followed predefined \js-specific dependency guidelines. Disagreements were resolved through discussion until consensus was reached.

An additional internal threat concerns incomplete dynamic analysis coverage. Since runtime dependency extraction depends on executing available test suites, dependencies in unexecuted code paths remain invisible to the dynamic approach. This may underestimate the effectiveness of runtime-based recommendation, particularly in systems with limited test coverage or highly input-sensitive behaviour.

To support reproducibility, we publicly release the implementation of \tool together with the benchmark metadata, experimental scripts, and evaluation artifacts used in this study at: \url{https://github.com/SEatSFU/caprese}.

%% file: Sections/RelatedWork.tex
\section{related work}

Static and dynamic approaches have been traditionally popular for change impact analysis \cite{PathImpact02003, Chianti2004, JRipples2005, EAS2005, progrmaSliceSize2007, huang2007precise, Impala2008, SDImpala2010, TaxonomyChange2010, ImpSlicing2011, coverateImpact2013, SENSAAnalysis2014, Diver2014, domSensitive2015, sunstatic2015, DIAPRO2016, jcia2018, KrishnaMurthy2018, allAppliableDependencies}.
%
%
     \textbf{Static impact analysis} techniques are usually based on static forward slicing of the programs
     and consider all possible executions
     \cite{theFirstImpactanalysis1996, programslicing1984, programslicing1988, AlgorithmicAnalysis2000}. 
     Therefore, the results tend to become very large and imprecise
     \cite{progrmaSliceSize2007, Impala2008, ImpSlicing2011}. 
     They are also not able to effectively support dynamic languages such as \js, where dynamism, event-driven execution,
     and asynchrony are commonly practiced
     \cite{AndreasenEsbenStatic, Sun2017, RichardsGregor, Alimadadi2016}. 
     \textbf{Dynamic analysis} approaches, on the other hand, rely on the dependency data gathered during executions
     \cite{dynamicSlicing1988, PathImpact02003, huang2007precise, SENSAAnalysis2014, DIAPRO2016}. 
     Despite being more precise, these techniques may suffer from incompleteness and high performance overheads.
     \cite{domSensitive2015, boostingChangePattern2021}. 
     Further, analysis of complex and heterogeneous platforms such as that of \js still remains challenging for
     dynamic analysis \cite{heterogeneous2011, boostingChangePattern2021}. 
     
     
        %

	Techniques such as \textbf{mining software repositories} and \textbf{information retrieval}
	have been widely used for impact analysis \cite{Zimmermann2004,  Zimmermann2005,  miningChangeRequests2005,  Canfora2006FineGI,  MiningSequencesKagdi2006,  Jashki2008, ImpactMiner2012, coChangePatternDetection2004, ImpRec2013, ImpRec2017, IntegratingConAndCoup2013, interactionAndHistory2014, TARMAQ2016, Rolfsnes2018, Islam2018, ATARI2020, IDCorrespondence2021}. 
        Prominent co-change detection tool ROSE uses association rules to realize evolutionary couplings. Rose takes a list of changed entities and recommends impacted files using a co-change detection algorithm \cite{Zimmermann2004, Zimmermann2005}. However, 
        ROSE works at the file level, and can not suggest any impacted files if there is no transaction containing all entities of the given change set. 
        \tarmaq resolves this issue \cite{TARMAQ2016} but still leads to imprecise and incomplete results 
        due to its limitation on the intersection between transactions and the change set and not including dynamic information, unlike \tool. 
    %
    Islam et al. \cite{Islam2018} considered the possibility that entities can be related even if they have not co-changed in the past. 
    To capture these couplings, they applied transitivity on the regular co-changes. 
    However, their technique is time-consuming in real-time usage and, based on their evaluation, results in lower precision than \tarmaq in some cases.
    In a recent study, Moonen et al. \cite{ATARI2020} proposed an adaptive approach that considers a dynamically selected set of commits instead of the entire history and can achieve the same level of accuracy as \tarmaq.
    %
    Unlike \tool, these techniques are unable to find previously-unknown patterns of various lengths with any common subset with the change set and are less precise. Further, \tool's dynamic analysis enhances the precision of our results considerably.

    Some have also investigated applying \textbf{sequential pattern mining} algorithms in order to
    find co-changes and their patterns in software history \cite{coChangePatternDetection2004, MiningSequencesKagdi2006, findGrainedKagdi2007}.
    Ying et al. \cite{coChangePatternDetection2004} first detect patterns with a frequency higher than a user-specific minimum support in the given change history. 
   Then, they report all of those patterns that include the file(s) in the user query. 
   However, their approach works at file level, and they evaluated their approach only for single-file change sets, showing the same performance as Rose.
    Kagdi et al. \cite{MiningSequencesKagdi2006} introduced sqminer. A co-change detection tool that uses SPADE, a sequential pattern mining algorithm, on data in the Subversion repository to find frequently occurring patterns with user-specific minimum support to detect co-changes at the file level. In a later study \cite{findGrainedKagdi2007}, they investigated the additional gain of using the same technique at a finer granularity than files. However, they did not publish any results.
    While useful, unlike \tool, none of these techniques 
    1) improves practicality by extracting only patterns that are relevant to the change set,
    2) is guided using ``confidence'' and thus works on small histories as well as large ones,
    3) ensures the fine level of granularity of \tool's analysis
    and 
    4) takes advantage of precise dynamic analysis.
    
     \textbf{Hybrid approaches} have gained in popularity for addressing the limitations of previous techniques.
	Blending static and dynamic analysis techniques can lead to improvement of precision and completeness
	of the findings
     \cite{Chianti2004, EAS2005, SDImpala2010, Diver2014, DiverSecondPaper2015, DiverOnline2018, KrishnaMurthy2018, SENSAAnalysis2014, domSensitive2015}.
     Algorithms based on information embed in software repository and information retrieval
      have shown to work well in conjunction with other analyses \cite{HassanHolt2004,  Hattori2008,  ImpactMiner2011, ImpactMiner2012, ImpactMinerTool2014, interactionAndHistory2014, CMsuggester2018, CMsuggester2018First, CMsuggester2020, IDCorrespondence2021}.
    In a recent study, Mondal et al. \cite{IDCorrespondence2021} proposed a complementary technique to the existing co-change detection techniques. A new metric called id-correspondence was suggested by the authors to complement the existing confidence and support measures. Their assumption was that if two co-changed entities are related, the identifiers existed in the modified lines of those entities have similar linguistic characteristics. The id-correspondence metric evaluates the degree of lexical similarity between the identifiers in the modified lines of the co-changed entities.
    Kagdi et al. \cite{IntegratingConAndCoup2013} investigated the combination of sqminer with information retrieval-based analysis of textual artifacts in a single software version. 
    These approaches did not consider dynamic dependencies, unlike \tool, while take into account the textual information. 

     Gethers et al. \cite{ImpactMiner2011, ImpactMiner2012} proposed an approach that takes a natural language user query and suggests an impact set using an adaptive combination of textual analysis and available contextual information, i.e., dynamic information manually collected by the user or functions co-changed with an initial software entity. One of the main differences between this approach and \tool is that their approach is based on textual similarities. They do not consider dynamic dependencies of a changed entity but narrow down the domain of search for finding functions textually similar to the user-given change request. Further, they used sqminer for co-change detection, which raises the same problems as stand-alone sqminer. 
     

     In a study conducted by Zanjani et al. \cite{interactionAndHistory2014}, the authors combined developers' interaction history in previous change requests
     with textual information extracted from source code and repository (e.g., source code, change request, and commit messages) 
     to suggest an impact-set for a given change-set.     
     However, this approach needs developers' interaction history that is not typically readily available.

    
    Some studies explored the relationships between the co-changed functions \cite{Mondal2013, UnknownChangePatterns2014, CMsuggester2018,  Silva2019, GraphBasedMiningWild2019, Jiang2020, MiningChangePatterns2020, boostingChangePattern2021, QuickFixesCodeRepo2021}. In these studies, a change pattern is defined based on the relationship between two or more functions. For example, adding a new function and modifying an existing function to invoke the new function is a common change pattern that occurs frequently during software evolution.
    %
     Silva et al. \cite{Silva2019} study six co-change patterns to investigate if co-change patterns exist across programming languages 
    and how different co-change patterns relate to rippling, activity density, ownership, and team diversity on clusters. 
    In a recent study, Huang et al. \cite{boostingChangePattern2021} proposed a boosting approach to enhance the ranking of the impacted entities. 
    They identified common coupling dependencies between the changed entity and impacted entities across various historical change sets. 
    Then, they used these common coupling dependencies to improve the ranking of the detected impacted entities that have those dependencies.
    %
    
    In research on bug fixes with multi-entry edits in Java programs, 
    Wang et al. \cite{CMsuggester2018} investigated whether there are repeated bug fixes that change multiple program entities.
    Then, inspired by this study, they developed CMSuggester---an approach that suggests complementary changes for multi-entity edits
    in Java applications that adds a field and modifies one or more methods to access the field \cite{CMsuggester2018, CMsuggester2018First, CMsuggester2020, GraphBasedChangePatterns2022}.
    Similar to this paper, Jiang et al. \cite{Jiang2020} proposed a machine learning based approach that finds all co-changed functions matching three pre-defined co-change patterns in software repository.  
    Unlike \tool, these approaches are limited to a pre-defined set of change patterns. They can only detect a potentially impacted function if it fits into their pre-defined categories of change. Further, they do not consider dynamic dependencies in the code.

%% file: Sections/Conclusion.tex
\section{Concluding Remarks}
\label{sec:conclusion}

Change impact recommendation is an important software engineering task that supports maintenance, debugging, regression testing, and code review. In highly dynamic languages such as \js, accurately identifying impacted code remains challenging due to implicit runtime dependencies arising from callbacks, asynchronous execution, events, and shared mutable state.

In this paper, we engineered and evaluated three practical function-level change impact recommendation techniques for \js: a repository-based approach, a dynamic dependency-based approach, and a hybrid approach that combines both signals. We evaluated these techniques on \numberOfSubjects\ open-source Node.js applications using expert-curated Reference Inspection Sets.

Our results show that repository-based and dynamic analyses capture complementary signals with surprisingly low overlap. Dynamic recommendation achieves the highest precision under small inspection budgets, while repository-based recommendation identifies additional historically coupled candidates under broader inspection budgets. The hybrid approach provides a balanced alternative by leveraging both runtime and historical information.

Overall, our findings suggest that no single technique sufficiently captures change impact in highly dynamic systems such as \js. Practical recommendation tools therefore benefit from combining complementary runtime and evolutionary signals to better support developer inspection and maintenance tasks.